%% file: SG.tex
\documentclass[sigconf, nonacm]{acmart}

\usepackage{caption}
\usepackage{multirow}
\usepackage{titlesec}
\usepackage{setspace}
\usepackage{url}
\usepackage{cleveref}
\usepackage{enumitem}
\usepackage[misc,geometry]{ifsym}
\usepackage{balance}

\makeatletter
\def\@ACM@checkaffil{% Only warnings
    \if@ACM@instpresent\else
    \ClassWarningNoLine{\@classname}{No institution present for an affiliation}%
    \fi
    \if@ACM@citypresent\else
    \ClassWarningNoLine{\@classname}{No city present for an affiliation}%
    \fi
    \if@ACM@countrypresent\else
        \ClassWarningNoLine{\@classname}{No country present for an affiliation}%
    \fi
}
\makeatother

\newcommand{\ie}{\textit{i.e.,} }
\newcommand{\eg}{\textit{e.g.,} }

\begin{document}

\title{The Software Genome Project: Venture to the Genomic Pathways of Open Source Software and Its Applications}

\author{Yueming Wu, Chengwei Liu$^*$, Yang Liu}
\affiliation{%
 \institution{Nanyang Technological University, Singapore}
}
\affiliation{%
  \institution{\{yueming.wu,\hspace{0.2em}chengwei.liu,\hspace{0.2em}yangliu\}@ntu.edu.sg}
}
\thanks{$*$ Chengwei Liu is the corresponding author.}

% \author{Yueming Wu}
% % \orcid{0009-0005-9832-8948}
% \email{yueming.wu@ntu.edu.sg}
% \affiliation{%
%   \institution{Nanyang Technological University}
%   \city{}
%   \country{Singapore}
% }

% \author{Chengwei Liu}
% % \orcid{0000-0003-1175-2753}
% \email{chengwei.liu@ntu.edu.sg}
% \affiliation{%
%   \institution{Nanyang Technological University}
%   \city{}
%   \country{Singapore}
% }

% \author{Yang Liu}
% % \orcid{0000-0001-7300-9215}
% \email{yangliu@ntu.edu.sg}
% \affiliation{%
%   \institution{Nanyang Technological University}
%   \city{}
%   \country{Singapore}
% }

\pagestyle{plain}

\input{outline/0_Abstract}

\maketitle

\input{outline/1_Introduction}
\input{outline/2_Terminology}
\input{outline/3_Genome_Construction}
\input{outline/4_Genome_Annotation}
\input{outline/5_Genome_Analysis}
\input{outline/6_Genome_based_Appilcations}
\input{outline/7_Open_Source_Governance}

\input{outline/8_Conclusion}

\balance
\bibliographystyle{ACM-Reference-Format}

\bibliography{SG}

\end{document}

%% file: outline/0_Abstract.tex
\begin{abstract}

With the boom in modern software development, open-source software has become an integral part of various industries, driving progress in computer science. 
However, the immense complexity and diversity of the open-source ecosystem also pose a series of challenges, including issues of quality, security, management, maintenance, compliance, and sustainability. 
Existing open-source governance approaches, while excelling in community building and collaboration, still face shortcomings in decentralized management, security, and maintenance. 
To address these challenges, inspired by the Human Genome Project, we treat the software source code as software DNA and propose the \textbf{Software Genome Project}, which is geared towards the secure monitoring and exploitation of open-source software. 
By identifying and labeling integrated and classified code features at a fine-grained level, and effectively identifying safeguards for functional implementations and non-functional requirements at different levels of granularity, Software Genome Project builds a complete set of software genome maps to help developers and managers gain a deeper understanding of software complexity and diversity. 
By dissecting and summarizing functional and undesirable genes, Software Genome Project helps facilitate targeted software remediation and optimization, provides valuable insight and understanding of the entire software ecosystem, and supports critical development tasks such as technology selection and open source governance. 
This project is expected to drive the evolution of software development towards more efficient, reliable, and sustainable software solutions.

\end{abstract}


%    <concept>
%        <concept_id>10011007.10011074.10011075.10011077</concept_id>
%        <concept_desc>Software and its engineering~Software design engineering</concept_desc>
%        <concept_significance>500</concept_significance>
%        </concept>
%  </ccs2012>

% \ccsdesc[500]{Software and its engineering~Software design engineering}

\keywords{Software Genome Project, Software Composition, Software Vulnerability, Third-party Library, Open Source Governance}

%% file: outline/1_Introduction.tex
\section{Introduction}
With the widespread adoption of open-source technology, open-source software projects have become a core infrastructure in modern software development. 
As the rapid growth of open-source ecosystems, the diversity and complexity of open-source codes also bring great energy to software ecosystem \cite{franco2017open}. 
However, due to the freedom, openness, and flexibility of open-source software, such popularity also brings many unignorable issues that could have significant repercussions, making the governance of open-source software a crucial concern \cite{prosandcons}. 
For instance, most of open-source projects is contributed for love, and there could be less regulated assurance and management during development, which makes the quality of open-source software varies. 
The security factor is another major concern of adopting open-source software in real-world development, the open collaboration in open-source projects make it easy to be maliciously compromised, such as planting backdoors and inserting malicious codes, and openly discussed vulnerabilities could also be utilized to compromise systems built on open-source software \cite{security2023}. 
Moreover, open-source projects could also suffer from lacking professional management, leading to insufficient supports and poor maintenance, such as missing necessary documentations and manuals, non-punctual reactions to critical issues, unclear legitimate licenses, leading to poor sustainability of open-source projects. 
These issues could significantly impede the adoption and management of open-source software in modern software development.

In response to the aforementioned challenges, we adopt a unique approach that views software code as analogous to biological DNA and  take a page from the Human Genome Project (HGP)'s playbook, leading to the inception of the Software Genome Project (SGP).
The HGP \cite{hgp, olson1993human}, launched in the early 1990s, stands as one of the most remarkable scientific endeavors in history. 
Its primary aim was to decode and comprehensively map all the genes of the human species. 
This monumental project sought to understand human biology at the molecular level, offering insights into our genetic heritage, hereditary diseases, and the fundamental building blocks of human life.
Taking a cue from the HGP, we introduce the SGP as a parallel initiative focusing on open source software. 
SGP's overarching goal is to achieve a comprehensive genetic mapping of software, akin to the HGP's sequencing of the human genome. 
In the realm of software, this translates to dissecting and annotating the complete genetic structure of code, from its most elementary components to its most intricate functionalities.
By extending the analogy between software code and biological DNA, SGP aspires to provide profound insights into open source software's genetic makeup. 
Such insights open the door to more effective governance, maintenance, and security in the world of software development.

The SGP, influenced by the transformative HGP, presents a comprehensive approach to revolutionize the open-source software landscape. 
Comprising five main phases, it harnesses the insights from human genetics to comprehend software code intricacies. 
Beginning with the construction of a comprehensive software genome and intelligence genome, this initiative paves the way for robust security analyses and defenses. 
After collecting the genome, we apply detailed annotation to discover software genes, the functionality of genes, and the functionality of intelligence genome. 
The subsequent phases delve deeper into genome analysis, offering insights into software gene relationships, families, and evolution. 
This knowledge is leveraged to facilitate comprehensive software composition analysis, akin to genetic testing, and software maintenance, similar to genetic engineering. 
Ultimately, the SGP serves as a catalyst for open source governance, fostering real-time security, compatibility, and compliance in the dynamic world of open source software.
Overall, the SGP, inspired by the concept of DNA in genetics, undertakes a systematic and comprehensive approach to enhance open-source software security. 
By treating software code as genetic material and adapting genetic principles, we can revolutionize the understanding, evaluation, and protection of open-source software. 

This paper makes the following contributions:
\begin{itemize}[leftmargin=8pt]
    \item \textbf{Proposal of Software Genome Project:} 
    We pioneer the establishment of a mapping correlation between software and DNA. 
    By this, we propose the first notion of software genome project. 
    \item \textbf{Construction of Software Gene Library:}
    Leveraging the innovative mapping approach, we create the first software genome and a groundbreaking software gene library.
    \item \textbf{Largest Software Intelligence Collection:} 
    We undertake the collection of the most extensive software intelligence, encompassing vulnerabilities, malicious code, and sensitive code. 
    \item \textbf{Innovative Software Genome Annotation:} 
    We design novel software gene analysis techniques, which are capable of delivering code details such as gene families and gene evolution.
    \item \textbf{Gene-based Software Composition Analysis:} 
    We offer inventive methods for analyzing software composition using genes, providing detailed insights into their utilization in software.
    \item \textbf{Gene-based Software Maintenance:} 
    We implement an original framework for software patching and refactoring, ensuring timely and effective software protection measures are in place.
    \item \textbf{Comprehensive Open Source Governance:} 
    We propose a comprehensive framework to facilitate real-time prewarning, package monitoring, compatible semantic versioning, and license compliance analysis.
\end{itemize}

\par \noindent \textbf{Paper organization.} The remainder of the paper is organized as follows.
Section 2 introduces the background and terminology used in our paper.
Section 3 shows the procedure of genome construction. 
Section 4 depicts the procedure of genome annotation. 
Section 5 describes the procedure of genome analysis. 
Section 6 represents some genome-based applications. 
Section 7 discusses the open source governance. 
Section 8 concludes the present paper.

%% file: outline/2_Terminology.tex
\section{Terminology}
In this section, we begin by introducing the Central Dogma of Molecular Biology and the Human Genome Project (HGP). 
Drawing inspiration from the HGP, we establish a mapping relationship between DNA and software, leading to the proposal of the Software Genome Project (SGP).

\subsection{Central Dogma of Molecular Biology}
The Central Dogma of Molecular Biology \cite{cobb201760, crick1958protein}, a fundamental concept in genetics, describes the flow of genetic information within a biological system. 
This process begins with the transcription of DNA into messenger RNA (mRNA). 
The mRNA then undergoes translation to form proteins, which are crucial for various cellular functions. 
This directional flow of information – from DNA to RNA to protein – is pivotal in understanding how genetic information is expressed in living organisms.
In a similar vein, in computer science, the transformation of source code into executable programs mirrors aspects of the Central Dogma. 
Source code, much like DNA, contains the original instructions or information. 
Compiler acts like mRNA, converting this source code into a executable program, just as mRNA translates genetic information into proteins.
This executable program is what computers execute to perform specific tasks. 
Additionally, the replication of source code to create clone code in software development is analogous to the replication of DNA in biological systems, where the genetic code is duplicated to ensure continuity and fidelity of information across generations. 
This parallel between the Central Dogma of Molecular Biology and computer programming highlights a fascinating intersection of concepts across biology and computer science.

\begin{figure}[ht]
\includegraphics[width=0.47\textwidth]{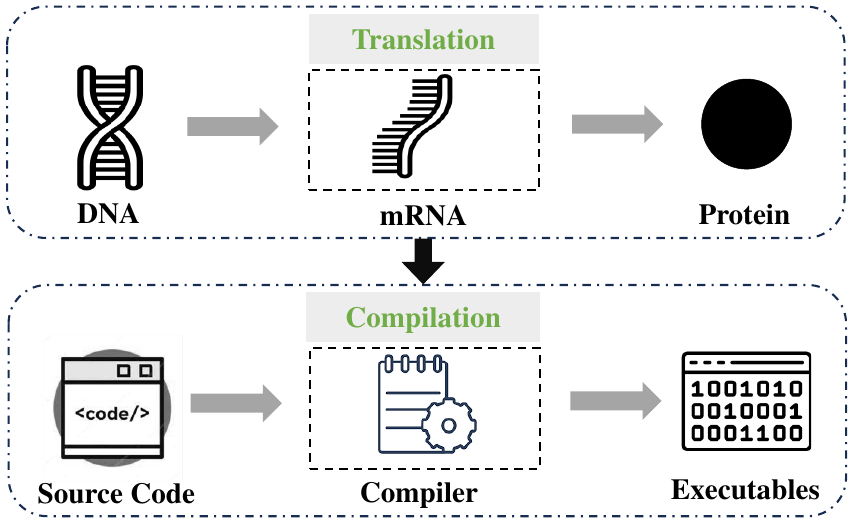}
\caption{Central dogma of molecular biology and the process of source code to executables}
\label{fig:central}
%\vspace{-1em}
\end{figure}

Based on the mapping relationship established by the Central Dogma of Molecular Biology, we can draw analogies from biology in three aspects as shwon in Figure \ref{fig:central}: source code to DNA, compiler to mRNA, and executable program to protein. 
In this paper, our primary focus is on the detailed analysis of source code (software DNA), with the objective of applying insights from the well-established field of DNA development to enhance open source governance.
In future work, we plan to delve into more detailed analyses of the other two aspects: compiler (software mRNA) and executable program (software protein). 
This comprehensive approach aims to leverage biological insights for advancing our understanding and practices in software development and governance.

\begin{figure*}[h]
\includegraphics[width=0.98\textwidth]{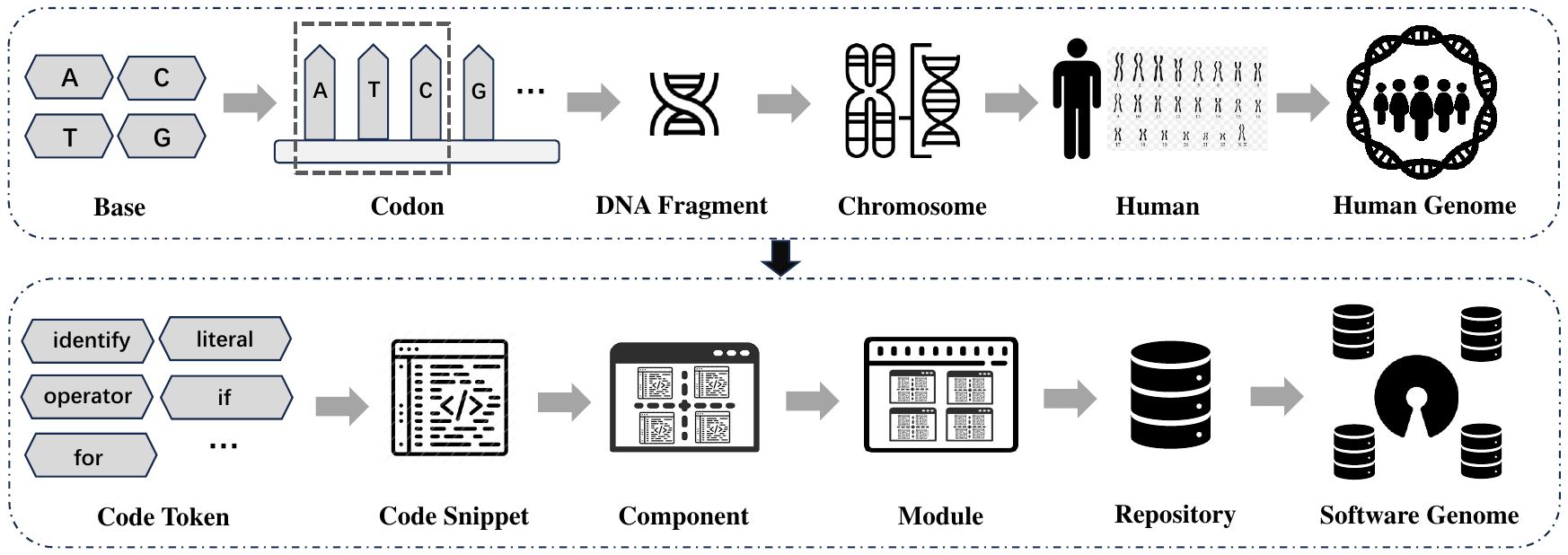}
\caption{Mapping relationship between DNA and open-source software}
\label{fig:mapping}
%\vspace{-1em}
\end{figure*}

% \subsection{Human Genome}
\subsection{Human Genome Project}
Before introducing software DNA, we first give a brief description of the Human Genome Project (HGP) \cite{hgp, olson1993human} and then clarify some terms used in Human Genome such as DNA.
The HGP  was a major global scientific effort initiated in 1987 and formally launched in 1990. 
It aimed to decode the 3 billion base pairs within human chromosomes, creating a comprehensive map of the human genome, including both coding and non-coding sequences. 
This collaborative endeavor, involving scientists from various nations, culminated in April 2003. 
The HGP is often considered one of the most significant scientific projects in history, akin to the Manhattan Atomic Bomb Project \cite{gosling1999manhattan} and the Apollo Project \cite{apollo}. 
It created a reference sequence for the human genome, facilitating research and deepening our understanding of human genetic diversity. 
The outcomes of the HGP are invaluable, serving as a foundational resource for medicine, biology, and genetics, aiding in the study of hereditary diseases, personalized medicine, pharmaceutical development, and gene-environment interactions.
The basic terms used in Human Genome \cite{venter2001sequence, pennisi2001human} are as follows:
\begin{itemize}[leftmargin=8pt]
    \item \textbf{Base}: 
    A base is a chemical unit that makes up DNA or RNA molecules. 
    In DNA, there are four types of bases: adenine (A), cytosine (C), guanine (G), and thymine (T). 
    They form base pairs (A-T and C-G) and are essential for encoding genetic information.
    \item \textbf{Codon}: 
    A codon is a sequence of three consecutive bases in DNA or RNA.
    It represents the basic unit of the genetic code, specifying an amino acid in protein synthesis.
    Codons guide protein synthesis by determining amino acid sequences, and there are 64 unique codons, including three stop codons that terminate protein synthesis. 
    \item \textbf{DNA Fragment}: 
    A DNA fragment is a part of the DNA molecule that can be a coding region or a non-coding region.
    A coding region is a gene that encodes proteins, which is essential component of an organism's functions.
    A non-coding region, while not directly encoding proteins, can also play a crucial role in gene regulation and other cellular processes.
    \item \textbf{Chromosome}: 
    A chromosome is a linear structure made of DNA and associated proteins. 
    It contains a long DNA molecule, tightly coiled to ensure orderly genetic information transmission.
    Chromosomes have an organized arrangement of genes and non-coding regions, ensuring accurate gene expression and inheritance.
    \item \textbf{Human}:
    A human typically contains 46 chromosomes, organized into 23 pairs.
    Males have one pair of sex chromosomes, comprising one X and one Y chromosome (XY), while females have one pair of X chromosomes (XX). 
    \item \textbf{Human Genome}:
    The human genome refers to the complete collection of genetic information within the human body. 
    The DNA differences between different individuals account for a relatively small portion of the entire human genome.
    The human genome is an extensive molecular database, consisting of billions of base pairs that form the genetic code of human beings.
\end{itemize}

In conclusion, the human genome, in its entirety, is the collective repository of all genetic information contained within human beings. 
This vast genetic archive includes not only the complete set of genes encoded in DNA, but also extends to non-coding regions, regulatory elements and additional DNA sequences that are intricately linked to heritable traits and biological functions. 
In essence, the genome serves as the genetic blueprint within an organism, meticulously dictating its growth, development, functionality and distinctive characteristics.
Genes, embedded within the genome, represent specific fragments of DNA with the essential function of encoding proteins. 
Genomic research is emerging as an indispensable endeavour, providing profound insights into the genetic make-up of organisms, evolution, health and disease. 
It reveals the intricate interplay between genes and how they collectively influence the characteristics and behaviour of an organism.

\subsection{Software Genome Project}

Inspired by human genome, we adopt a perspective that treats software code as akin to biological DNA. 
In this part, we first introduce several terms used in software genome.
\begin{itemize}[leftmargin=8pt]
    \item \textbf{Code Token}:
    A code token is the fundamental building block of software, similar to how a base is a fundamental unit in DNA. 
    Code tokens include elements like ``Identifier'', ``Keywords'',`` Modifier'', and ``Operator'', and they form the basic syntax of a programming language.
    For instance, there are 15 token types in Java programming language \cite{wu2022detecting}.
    \item \textbf{Code Snippet}:
    A code snippet is a sequence of code tokens, just as a codon is a sequence of three consecutive bases. 
    Code snippets represent a basic unit of the software's functionality, specifying a specific operation or function in the program.
    \item \textbf{Component}:
    A component in software is a larger unit that comprises multiple code snippets. 
    It can include functionalities such as functions, classes, or files. 
    Components can be either coding regions, which directly contribute to the software's functionalities, or non-coding regions that support software maintenance and operation.
    \item \textbf{Module}:
    A module in software is an organized structure that contains multiple components for organizing and isolating functionalities.
    Modules aim to enhance code organization, promote reusability, and facilitate maintenance in software projects, akin to chromosomes in genetics, which help maintain order and transmit genetic information.
    \item \textbf{Repository}:
    A repository, similar to a human being, is a complete entity. It encompasses all the modules, components, and code snippets that make up the functional aspects of the software. 
    A repository is designed to serve specific purposes and functions, just as a human being has a specific genetic makeup and characteristics.
    \item \textbf{Software Genome}:
    The software genome is a metaphorical term used to describe the complete collection of code within a software system. 
    It serves as a comprehensive reference of software code, akin to the human genome.
\end{itemize}

To give a better illustration, we use Figure \ref{fig:mapping} to show the mapping relationship between DNA and software.
Through the figure, we see that an individual carries multiple chromosome, each comprising coding and non-coding segments. 
Segments of DNA that encode proteins are referred to as genes. 
A DNA fragment consists of multiple codons, with each codon made up of three bases. 
Looking at software architecture, a software repository is a composition of various modules, each module housing several components. 
Components, in turn, contain multiple code snippets, and each snippet's purpose is achieved through different code token combinations. 
When viewing a software as a human, all of its code constitutes its genetic material, and the collective code of all open-source software in the open source community constitutes what we refer to as the software genome.
Building on this analogy, we introduce the concept of the Software Genome Project (SGP), aiming to apply a HGP-like approach to the software community.

\begin{figure}[h]
\includegraphics[width=0.47\textwidth]{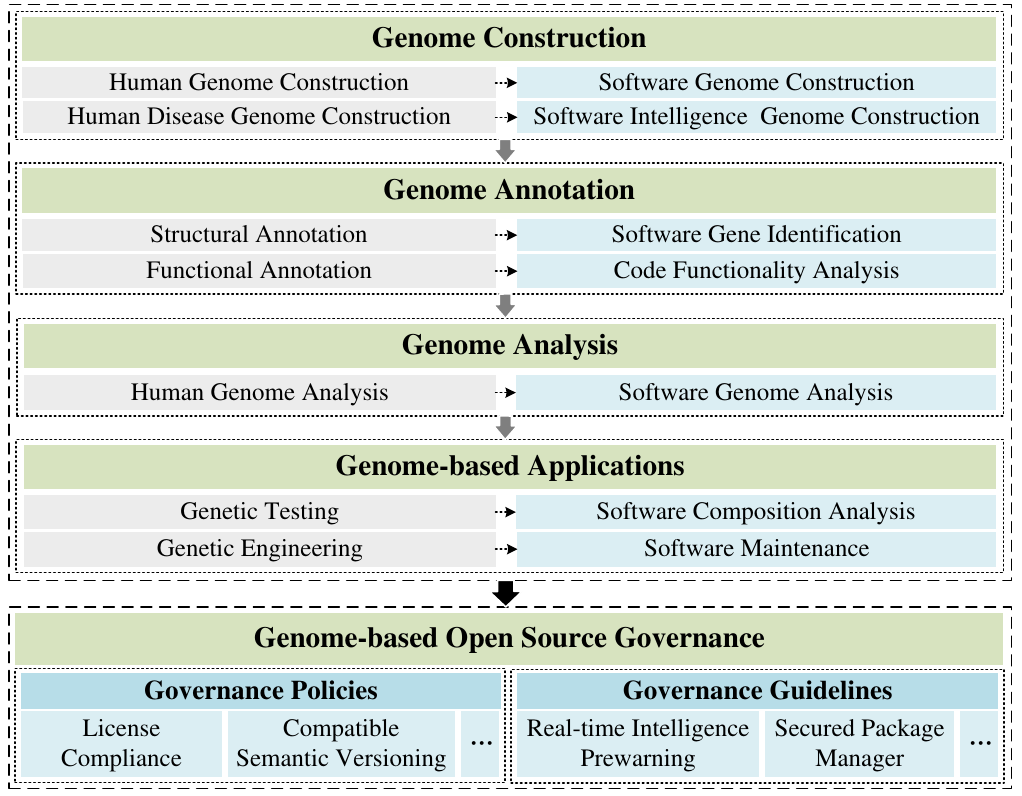}
\caption{Framework of software genome project}
\label{fig:sgp}
%\vspace{-1em}
\end{figure}

As depicted in Figure \ref{fig:sgp}, our SGP comprises five phases: \emph{Genome Construction}, \emph{Genome Annotation}, \emph{Genome Analysis}, \emph{Genome-based Applications}, and \emph{Genome-based Open Source Governance}. 
The scope of this initiative extends beyond the mere construction of the software genome, it also involves the assembly of a software intelligence genome, drawing parallels with the human disease genome. 
This comprehensive dataset equips us to conduct exhaustive security analyses and enact robust software defenses.
Subsequently, we delve into structural annotation to identify software genes that hold paramount significance within the open-source community, meriting ongoing maintenance. 
Moreover, we also perform functional annotation to enable us to comprehend the roles and functionalities of all software genes and the software intelligence genome. 
Armed with this knowledge, we proceed to genome analysis, delving into aspects such as dependency, clustering, portrait, and evolution. 
The knowledge acquired through genome analysis can enhance the effectiveness of software composition analysis and software maintenance, akin to the principles of genetic testing and genetic engineering.
Moreover, unlike the natural evolution of biological genes, we got the chance to participate and guide the evolution of software ecosystems. 
Our ultimate goal of SGP is to chart a course for open-source governance, ensuring the development of a robust, real-time, secured, and effective open-source community.

%% file: outline/3_Genome_Construction.tex
\section{Genome Construction}
\begin{figure*}[h]
\includegraphics[width=1\textwidth]{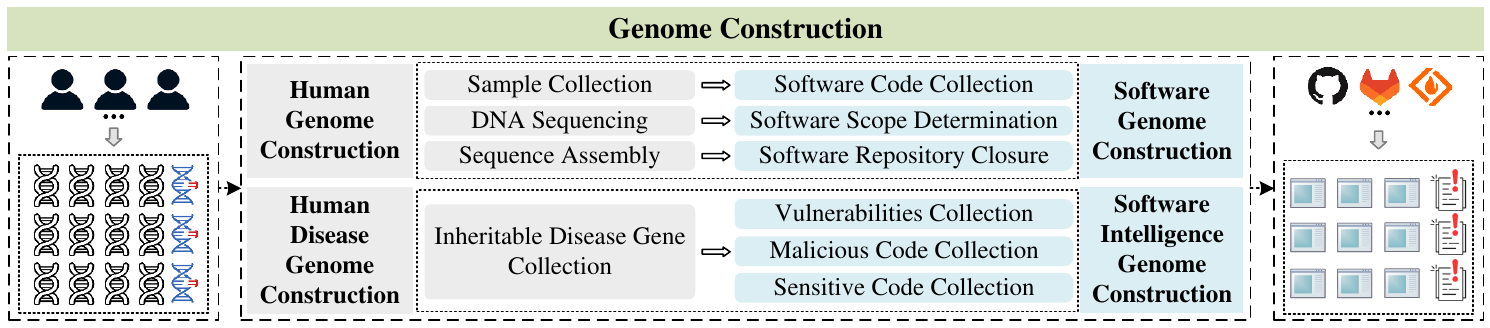}
\caption{Genome construction of software genome project}
\label{fig:genome-construction}
%\vspace{-1em}
\end{figure*}

This phase aims to lay the foundation for creating the world's inaugural software genome and, subsequently, to employ it as the basis for crafting an innovative software gene repository. \Cref{fig:genome-construction} depicts the overview of genome construction, including the software genome and software intelligence genome that are hidden in the software genes.

\subsection{Software Genome Construction}

\subsubsection{Sample Collection → Software Code Collection.}\label{subsubsec:collect}
% Given the excessive amount of software repositories in real world, it is crucial to fully examine the consistency of all software within reasonable complexity and resource consuming. 
% In this task, we aim to propose a systematic solution on the efficient collection of the most valuable repositories from the perspective of constructing software genome. 
% Therefore, we initialize the construction of software genome by sorting out the most recognized software repositories. 
% Considering that most existing research on software repository mining have proposed various criteria on selecting representative repositories for experiment and analysis. 
% We first revisit these papers to identify the mostly recognized metrics on selecting representative repositories, such as stars, forks, issues, commits and contributors, etc. 
% Specifically, we investigate both their representativeness of important software repositories and availability from the aspect of ecosystem-wide acquisition, so that they can be properly acquired or calculated for large scale analysis.
% Based on these, we interpret their distribution in the entire open-source ecosystem and set the threshold as a coarse-grained bar to sort out the most recognized repositories as the initial candidate projects. 
% Specifically, as most open-source repositories are with limited value (\ie not well organized and maintained), we aim to set the threshold as a relatively low bar to exclude meaningless repositories based on distribution analysis. 

Given the proliferation of software repositories in real-world software repositories, it becomes imperative to rigorously collect software samples that can cover the most software genes while ensuring computational efficiency. Therefore, prioritizing software repositories so that those most prominent software repositories that provide most software genes can be prioritized to be sampled is essential for software genome construction.
% In this study, our objective is to introduce a methodical approach for the effective aggregation of repositories that hold significant value in the context of constructing a software genome. 
% To initiate the construction of this genome, we prioritize the identification and categorization of the most prominent software repositories. 
Notably, a plethora of existing studies~\cite{hu2016influence, 7816479, zerouali2019diversity} on software repository mining have delineated various criteria for selecting emblematic repositories for empirical analysis. We commence by reviewing these scholarly works to discern the most frequently cited metrics for repository selection, including but not limited to stars, forks, issues, commits, and contributors. Our investigation encompasses an evaluation of both the representativeness of pivotal software repositories and their accessibility for ecosystem-wide data extraction, ensuring that these metrics can be efficiently procured or computed for expansive analysis. Drawing on these findings, preliminary thresholds can be established to identify the most notable repositories for potential inclusion. Given that a substantial portion of open-source repositories may lack intrinsic value (often due to sub-optimal organization or maintenance), we strategically set this threshold to exclude repositories based on our distributional analysis.

\subsubsection{DNA Sequencing → Software Scope Determination.}\label{subsubsec:scope}

Subsequently, we extract the source code from each shortlisted repository to discern the characteristics imparted by distinct code segments. Specifically, from a genomic diversity standpoint, we appraise a software repository based on its contribution to code heterogeneity within the ecosystem. Theoretically, a repository enriched with unique and emblematic code segments is posited to offer greater value compared to repositories predominantly comprised of replicated content with fewer distinctive code fragments. Hence, predicated on the chosen candidate repositories, we initially quantify their code snippet level significance by aggregating their distinctiveness. Specifically, the code snippet consists of sequential lines of code, which should be meaningful of program functionalities. Therefore, we fuse different metrics, such as Halstead Volume~\cite{Halstead}, Cyclomatic Complexity~\cite{Cyclomatic}, and lines of codes (LOC), to reflect the importance of software genes of code snippets, and based on this, the contribution of software repositories can also be measured. Considering same code snippets could be from different repositories, we share the contributed value of code snippets when measuring the contribution of different repositories.

\subsubsection{Sequence Assembly → Software Repository Closure.}

While \Cref{subsubsec:collect} identifies prominent repositories for software genome extraction, some may be overlooked due to specific selection criteria. To rectify this, we propose a bi-directional strategy to incrementally integrate emblematic repositories and exclude anomalous ones with limited value. Given that many repositories might be eliminated by our criteria, indiscriminate traversal could be inefficient. Therefore, we aim to devise an algorithm to prioritize the reintroduction of initially omitted repositories, streamlining the software genome's completion.

We categorize repositories likely to be prominent, organizing them to augment the identified software genes in \Cref{subsubsec:scope}. Each repository is characterized by a vector of metrics from \Cref{subsubsec:collect}. Using the Local Outliers Factor (LOF)~\cite{Lof}, we identify potentially erroneously excluded projects. Repositories deemed outliers are believed to possess exceptional value compared to underperforming ones. For excluded repositories, we rank them using LOF and review them to enhance the construction of the software genome until reaching a point of diminishing cost-effectiveness.

\subsection{Software Intelligence Genome Construction}
Generally, we consider three major types of software intelligence that could threaten software if they are inherited, security vulnerability, malicious code, and sensitive code snippets.

\subsubsection{Inheritable Disease Gene Collection → Vulnerability Collection.}

% Collecting software vulnerability data is essential for effective risk management, informed decision-making, and maintaining the security of software systems. 
% This data enables organizations to prioritize and apply security patches, mitigate potential risks, ensure regulatory compliance, and enhance incident response strategies. 
% It empowers security teams with threat intelligence, aids in vendor assessment, supports security training efforts, and fosters a culture of continuous improvement. 
% Organizations can proactively protect against emerging threats by analyzing vulnerability data, making strategic software choices, and demonstrating their commitment to robust cybersecurity practices. 
% Our overarching aim is to establish a comprehensive data platform housing curated, current, and high-caliber vulnerability data. 
% This data reservoir will serve as the foundation for the creation of a secure package manager and an online alarm system, enabling enhanced software security measures.

As one of the most critical threats to software security, software vulnerability is pivotal for risk management, decision-making, and software security, especially under the prevalence of reusing open-source software, which makes vulnerabilities inheritable to downstream users. 
Our primary goal is to develop a comprehensive platform containing up-to-date, high-quality vulnerability data, laying the groundwork for the open-source intelligence platform for secured software development and open-source governance.

\emph{\textbf{Vulnerability Data Collection and Evaluation.}}
% In the practical realm, a significant portion of vulnerability data originates from online databases such as CVE and NVD, each hosted by distinct entities with specific objectives. For instance, NVD, overseen by the U.S. government, strives to comprehensively document the landscape of existing vulnerabilities, whereas OSV, sponsored by Google, focuses exclusively on vulnerabilities present in widely utilized open-source libraries.
% Our objective is to examine each source of vulnerability data and assess its quality. To achieve this, we will establish a robust data pipeline to aggregate vulnerability data from diverse sources, implementing an incremental update approach to ensure the continuous accuracy of the data. Furthermore, we will develop a data dependency analysis framework to elucidate the provenance of the data and effectively eliminate any redundancies. Through attribution of individual vulnerability facets, we will undertake differential analyses to discern disparities in the portrayal of identical vulnerabilities.
Much vulnerability data are indexed from different online databases like CVE~\cite{cve} and NVD~\cite{cwe}, while each database may involve different entities for various purposes. For example, NVD aims to index all vulnerabilities, while Google's OSV~\cite{OSV} focuses on those in popular open-source libraries, and lays more effort on accuracy. Our goal is to evaluate the quality of data from each source and ensure high coverage and accuracy for vulnerability collection. First, a data pipeline is set up to collect vulnerability data from various sources incrementally. Next, a data dependency analysis framework is established to trace data origins and identify accurate data. By attributing specific vulnerability aspects, differences among data collected from different sources are analyzed and merged to construct a comprehensive vulnerability intelligence platform.

\emph{\textbf{Vulnerability Data Enhancement.}} 
% The vulnerability information procured from online sources may exhibit inaccuracies or omissions. Our objective is to harness advanced programming analysis techniques in conjunction with the capabilities of a large language model (LLM) to enhance the information. A case in point pertains to the established methodologies for ascertaining the scope of vulnerability impact, often reliant on human auditing, a process susceptible to errors. We intend to extrapolate this scope directly from the code lines. In the case of each CVE entry, we will perform an execution trace traversal of functions using program slicing techniques, outlining vulnerability constraints through symbolic execution. Subsequently, we will employ fuzzy matching techniques to align these constraints with susceptible lines of code, enabling rapid and precise identification of the vulnerability's presence within any given software iteration. Furthermore, we will train the language model to comprehend source code commits, thus enabling it to predict accurate commit messages and vulnerability descriptions. 
Vulnerability data from online sources could be incomplete. Traditional methods to determine vulnerability impact often depend on human auditing, which can be error-prone and miss key information~\cite{10.1145/3498537}. Instead, We adopt advanced programming analysis and a large language model (LLM) to derive detailed vulnerability data directly from code lines. For each CVE entry, we trace function executions using program slicing and define vulnerability constraints via symbolic execution. Based on this, fuzzy matching would be conducted to correlate these constraints with affected code lines, allowing for quick and accurate vulnerability identification in software versions. Additionally, LLMs would be also involved to understanding source code commits, aiding in predicting precise commit messages and vulnerability descriptions.

% This endeavor will significantly refine critical facets of vulnerability information, encompassing the scope of impact, patch information, and proof of concept (PoC) elucidations. All the downstream tasks such as vulnerability prediction, patch generation, and PoC detection will benefit from the improvement of the data quality.

\subsubsection{Inheritable Disease Gene Collection → Malicious Code Collection.}

\begin{figure*}[h]
\includegraphics[width=1\textwidth]{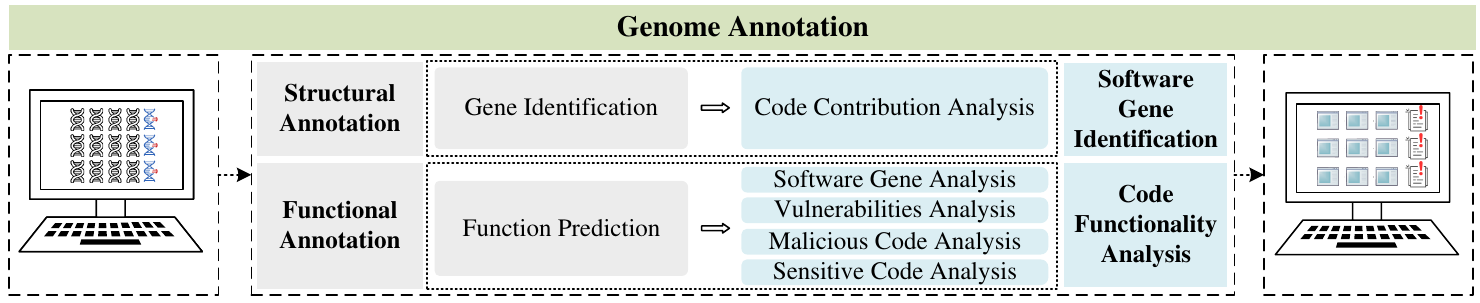}
\caption{Genome annotation of software genome project}
\label{fig:genome-annotation}
%\vspace{-1em}
\end{figure*}

Malicious code, manifesting as viruses, worms, and spyware, poses significant threats to software ecosystems by stealing data, hijacking systems, and causing financial and reputational harm. Researching malicious code is vital to understand evolving threats, develop detection techniques, and bolster defenses. We aim to compile a comprehensive malicious code dataset, incorporating source code and binary files from languages like Python, C, and Java. This dataset will deepen our understanding of malicious code characteristics, serving as a foundation for software genomics studies. We'll source data from various channels, such as underground forums, community platforms, and databases like VirusShare~\cite{VirusShare} and VX Heaven~\cite{vxheaven}, and will also mine GitHub for diverse code samples.

During data collection, we adapted strategies to each source's unique attributes. We developed web crawling modules for forums and platforms to automatically fetch malicious code-related content using specific keywords. For databases like VirusShare and VX Heaven, we leveraged API interfaces to retrieve data, including essential metadata and hash values. We also crafted modules to gather data from GitHub, focusing on multiple programming languages. To ensure data quality, we used hash algorithms for deduplication and performed integrity checks. The final dataset, annotated with details like origin, timestamp, and malicious code family, aims to be a comprehensive resource for future malicious code research.

\subsubsection{Inheritable Disease Gene Collection → Sensitive Code Collection.}

The open-source ecosystem, built on collaboration and shared knowledge, can unintentionally propagate sensitive or harmful content. Given the diverse backgrounds of its stakeholders, sensitive code in open-source repositories can lead to misunderstandings and misuse. The lack of a robust framework to manage sensitive content jeopardizes the safety, inclusivity, and ethical stance of open-source software. It's imperative to have clear guidelines to handle sensitive content, ensuring open source remains a constructive and secure domain for all. This is vital not only ethically but also to maintain trust in the open-source community.

Therefore, we aim to curate a dataset of sensitive code from open-source repositories. We focus on three types of sensitive code:
1) User-sensitive code: New contributors might inadvertently commit private data, like access tokens or database credentials. Such oversights can be exploited by malicious entities.
2) Behaviour-sensitive code: Some open-source components, while not malicious, require special system privileges. Attackers can misuse these privileges, leading to security breaches when these components are integrated into larger systems.
3) Politics-sensitive code: Open-source platforms, due to their global reach, can unintentionally host political biases or propaganda. Monitoring for political sensitivities is crucial to maintain project neutrality and prevent potential legal or ethical issues. This ensures open-source remains focused on technology, free from divisive political content.

% Collecting and monitoring these sensitive codes could benefit the proper detection and governance of open-source ecosystems, and analysing these sensitive codes could also strengthen the understandings of sensitive information existing in open-source repositories and promoting the detection of sensitive code from user aspect.

%% file: outline/4_Genome_Annotation.tex
\section{Genome Annotation}

After constructing the software genome, we perform a comprehensive annotation to discover ``software genes'' and their associated functionality.
Meanwhile, we also annotate the functionality of our software intelligence genome including vulnerabilities, malicious code, and sensitive code.
The detailed steps are shown in Figure \ref{fig:genome-annotation}.

\subsection{Software Gene Identification}
\subsubsection{Gene Identification → Code Contribution Analysis.}
Given a software genome, we employ multifaceted analyses to understand the structural properties and identify key software genes. 
Specifically, we first construct a call graph to represent the complex caller-callee relationships between all functions. 
Next, we conduct advanced social network centrality analysis such as degree centrality analysis \cite{freeman1978centrality}, closeness centrality analysis \cite{freeman1978centrality}, and betweenness centrality analysis \cite{freeman1977set} to identify key nodes (similar to``influencer'') that control the behaviour and information flow of the software.
Apart from analyzing the connections between functions, we also perform a comprehensive code analysis for each function to reveal the code complexity \cite{codecomplexity}.
Our previous work \cite{wu2023ossfp} has revealed that useless code always exhibits low code complexity because they do not implement core logic of the program.
Combining these local (\ie code complexity) and global (\ie centrality) insights, we create a ranking for all functions, highlighting their relative importance and contribution to the software ecosystem. 
This procedure blends complex analyses ranging from call graph construction to centrality assessment and code complexity assessment, culminating in a robust software gene pool that captures the genetic nature of software.

\subsection{Code Functionality Analysis}
After identifying all software genes, we'd like to assess their functionality one by one.
Furthermore, we also perform function prediction on the software intelligence genome to document their behaviors, potential impacts, and the context in which they can be exploited or pose risks.

\subsubsection{Function Prediction → Software Gene Analysis.}
As for software gene analysis, we begin by examining the code's structure, variables, and other program details to glean insights into its intended purpose.
We also explore any accompanying comments provided by developers during the code's creation, which can often offer invaluable insights into the code's intended functionality.
Moreover, if available, we will execute the code in a controlled environment by using some testing cases to verify that the code behaves as intended and produces the expected results. 
Our purpose is to answer fundamental questions about what the code achieves, how it accomplishes its goals, and the specific problem it addresses.
Once we have a solid understanding of the code's functionality, we proceed to provide essential code details for accurate documentation.
This document will serve as the dictionary of our software genes.

\subsubsection{Function Prediction → Vulnerabilities Analysis.}
As for vulnerabilities analysis, we begin by understanding the nature of the vulnerability. 
This includes identifying the specific type (\ie CWE type \cite{cwe}) and how it affects the software. 
In detail, we will determine how it can be triggered and what it may lead to in terms of unauthorized access, data breaches, or system compromises. 
Furthermore, we’d like to annotate the context in which the vulnerability can be exploited and describe the impact on the software such as the potential consequences of exploitation.
Finally, we will analyze potential scenarios where the vulnerability might be exploited, including attack vectors, attacker profiles, and the methods they might employ. 
This will help in assessing real-world risks.

\subsubsection{Function Prediction → Malicious Code Analysis.}
As for malicious code analysis, we begin by understanding the behavior such as its purpose (\eg data exfiltration \cite{dataexfiltration}) and how it operates within the software. 
Then, we will document its specific actions, payloads, and describe the conditions under which the malicious code is triggered or executed. 
This includes events, user actions, or system states that activate the malicious behavior. 
Also, we will analyze the payload carried by the malicious code and specify the types of malicious files, commands, or data manipulations it performs.
If the malicious code can propagate within the software or to other systems, we will also detail how this occurs. 

\subsubsection{Function Prediction → Sensitive Code Analysis.}
As for sensitive code analysis, we begin by identifying code segments that handle sensitive data such as user credentials, personal information, or financial data. 
Specifically, we describe how the sensitive data is processed within the code including data collection, storage, transmission, and disposal practices. 
In addition, we will document the access control mechanisms in place for sensitive data and detail who has permissions to interact with this data and under what conditions. 
Meanwhile, we consider any industry-specific compliance requirements (\eg GDPR \cite{gdpr}) or security standards that the code must adhere to, and annotate how the code complies with them.

By conducting functional annotation for software genes, vulnerabilities, malicious code, and sensitive code, we can provide essential documentation to guide developers, security experts, and users in understanding software and addressing potential risks and security issues.

%% file: outline/5_Genome_Analysis.tex
\section{Genome Analysis}

\begin{figure*}[h]
\includegraphics[width=01\textwidth]{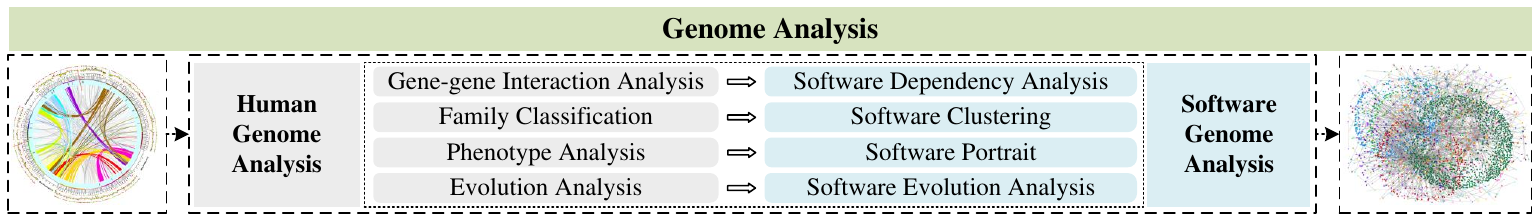}
\caption{Genome analysis of software genome project}
\label{fig:genome-analysis}
%\vspace{-1em}
\end{figure*}

% After the annotation of software genes, we perform in-depth analysis on our collected genomics, including gene-gene interactions analysis, gene family classification, gene phenotype analysis, and gene evolution analysis. They enable us to draw connections between individual genes, identify shared functionalities, and recognize dependencies that contribute to the software's overall functionality, quality, and evolution.

After annotating software genes, we conduct a thorough analysis of our genomics. This includes studying gene-gene interactions, classifying gene families, examining gene phenotypes, and gene evolution analysis. These insights help us understand gene interrelations, shared functionalities, and dependencies that influence the software's functionality and evolution.

\subsection{Software Genome Analysis}

\subsubsection{Gen-gene Interactions Analysis → Software Dependency Analysis.}

% Abundant evidence and accumulated experience show that the traits of certain complex diseases are collectively influenced by interactions among multiple genes. Consequently, investigating the interplay between genes from a biological perspective holds substantial value. In pursuit of comprehensive open-source intelligence analysis, we also extend our examination to the interactions among software genes, analogous to the dependencies within genes. Our analysis encompasses not only the directly defined software dependencies within package managers but also delves into more intricate relationships induced by code cloning. 

Ample evidence suggests that traits of complex diseases are shaped by interactions among multiple genes. Thus, exploring gene interplay is biologically significant. Similarly, in our open-source intelligence analysis, we probe into interactions among software genes, akin to gene dependencies. Our study covers not just direct software dependencies from package managers but also intricate relationships stemming from code cloning.

\emph{\textbf{Package Manager-based (PM-based) Dependency Analysis.}}
PM-based dependencies represent explicit relationships within a codebase, ensuring the integration and compatibility of external libraries or components. Defined within the package management system, they allow developers to seamlessly incorporate external resources. This structured approach lets software modules depend on external packages' functionalities. The package manager oversees the retrieval, installation, and versioning of these dependencies, ensuring systematic integration. Developers can specify package versions, ensuring consistent software behavior and compatibility. To extract these dependencies, we follow:
1) Analyze Package Files: Identify package management files, e.g., package.json for JavaScript or requirements.txt for Python.
2) Parse Dependency Definitions: Extract package names and versions from these files, indicating required external components.
3) Retrieve Metadata: Use the package manager's API to gather metadata about each dependency, including version and other dependencies.
4) Construct Dependency Graph: Link the project's modules to their respective external packages, including version constraints.
5) Analyze Interactions: Determine how dependencies function within the codebase, identifying their usage and contribution to the software.

\emph{\textbf{Code Clone-based (CC-based) Dependency Analysis.}}
% CC-based dependencies stem from the reuse of existing code fragments or patterns within the software. This form of dependency emerges when portions of code are duplicated across different parts of the software, often to replicate a specific functionality or design approach. While not explicitly indicated as dependencies within the codebase, these cloned code segments inherently create interconnections among various sections of the software. To extract CC-based dependencies from a software project, we will follow these steps: 1) Identify Cloned Fragments: Design novel code analysis tools to identify duplicated code fragments within the project. These tools can detect similar or identical code patterns across different parts of the codebase. 2) Trace Clones: Trace the instances of cloned code to understand where they are duplicated and how they are utilized across the software. 3) Analyze Impact: Analyze the impact of code changes in one cloned instance on other instances to understand how changes propagate through the codebase due to these clones. 4) Quantify Relationships: Quantify the relationships between cloned code fragments and identify shared functionalities, data structures, and patterns that contribute to software behavior. 5) Visualize Dependencies: Create visualizations or graphs that illustrate the connections between different cloned instances. This can help developers understand the intricate relationships and identify potential areas for optimization.
CC-based dependencies arise from code reuse within software. These dependencies manifest when code segments are duplicated, often to emulate specific functionalities. While not overtly labeled as dependencies, these code clones inherently link various software sections. To extract these dependencies, we follow: 1) Identify Cloned Fragments: Use advanced code analysis tools to detect duplicated code patterns throughout the project.
2) Trace Clones: Track cloned code instances to discern their distribution and application across the software.
3) Analyze Impact: Assess how changes in one clone affect others, understanding change propagation due to these clones.
4) Quantify Relationships: Measure the ties between clones, pinpointing shared functionalities and patterns influencing software behavior.
5) Visualize Dependencies: Generate visual representations of connections between clones, aiding developers in grasping these relationships and spotting optimization opportunities.

% Both types of dependencies contribute to the overall structure and behavior of the software. Package manager-based dependencies provide controlled integration with external resources, enhancing modularity and code reusability. Code clone-based dependencies, while not as overt, underscore the importance of code organization and shared patterns within the software. By considering both forms of dependency, a comprehensive view of the software's interconnectedness emerges, enabling developers to optimize performance, mitigate risks, and enhance software quality.

\subsubsection{Family Classification → Software Clustering.}

% The excessive adoption of open-source artifacts in modern software development has promoted the prosperity of open-source ecosystems, including both the published third-party libraries and code snippets that are widely reused by developers (i.e., software genes). With this, users can bypass the need to code from scratch and accomplish more complex and challenging tasks in easier ways. However, the vast and complex ecosystem of open-source artifacts make it difficult to properly comb the vast knowledge and coding practices, which makes it insufficient to reuse existing artifacts to accelerate modern software development (i.e., guiding LLMs for SE). In this case, we aim to dive into the applications of software genes and comb their usage practices and diversity in different domains.
% Our objective is to construct a comprehensive map to figure out the ability and diversity of software genes, including the classifications of software genes, their scope and applicable domains, usage practices, possible combinations, and their evolutions. Specifically, due to the diversity of software genes, they could be with different size and formats, we aim to dig out the possible combination of software genes in a mixed solution.

The rise of open-source components in software development has enriched open-source ecosystems, encompassing third-party libraries and frequently reused code, i.e., software genes. This allows developers to efficiently tackle complex tasks. However, navigating this vast ecosystem to optimize the reuse of artifacts, such as guiding LLMs for SE, remains challenging. Our focus is on delving into software genes' applications and understanding their diverse usage across domains.
Therefore, we aim to craft a detailed map of software genes' capabilities and diversity. This involves classifying them, pinpointing their domains, analyzing usage patterns, exploring potential combinations, and observing their evolution. Given their varied sizes and formats, we also intend to uncover hybrid combinations of these genes.

\emph{\textbf{Software Gene Clustering.}}
% First, the construction of software genome has unveiled the existence of software genes in different software repositories, including both open-source applications and third-party libraries, and we are able to utilize such existence to cluster software genes by the domains of applications and third-party libraries. Specifically, we plan to use two types of information to grouping software genes. 1) Co-existence. Software genes could be integrated into applications or third-party libraries within similar domains. In this case, we investigate the co-existence of software genes in open-source artifacts. If such co-existence is with high possibility, it may indicate that these software genes are usually used in software within similar domains. 2) Dependency Relationship. As the most important part of open-source ecosystems, third-party libraries play important roles in software development, which also provide significant portion of software genes. Therefore, we trace the dependency relationships among third party libraries with abundant genes. Considering that dependency relationships naturally indicate the connection between software artifacts, software genes that are within third-party libraries on the same dependency paths should also naturally have implicit dependencies in specific domains. Therefore, we aim to combine this information as the distance factor to classify software genes to different domains.
The development of the software genome has revealed the presence of software genes across various software repositories, encompassing both open-source applications and third-party libraries. Leveraging this, we aim to cluster software genes based on application and library domains. Our clustering approach hinges on two key information types:
1) Co-existence: Software genes often integrate into applications or libraries within similar domains. We examine the co-occurrence of software genes in open-source artifacts. A high co-existence probability suggests these genes typically feature in software from similar domains.
2) Dependency Relationship: Third-party libraries, pivotal in software development, contribute significantly to software genes. We map the dependency relationships among gene-rich third-party libraries. Given that these relationships inherently signify connections between software artifacts, genes within libraries sharing dependency paths likely have implicit domain-specific dependencies. We'll use this relationship data as a distance metric to categorize software genes by domain.

\emph{\textbf{Software Gene Semantics.}}
% After clustering these software genes, we aim to comb these clustered software genes and identify the semantics of them. Specifically, we identify the correlation of software genes by mining the frequent patterns appearing in existing software, based on which, we aggregate these genes to identify their usage and semantics by integrating LLM techniques. Moreover, with the classification of software domains, we can interpret software genes with their usage and presentations on software functionalities. Furthermore, based on their usage, the critical genes for different software domains can be also identified, such as the common practices and fundamental software genes, which could be further utilized to guide future software development to improve efficiency.
After clustering the software genes, our next step is to discern their semantics. We determine the correlation between software genes by extracting recurrent patterns in existing software. Leveraging LLM techniques, we then aggregate these genes to understand their usage and semantics. With the software domain classifications in place, we can contextualize software genes in terms of their functional contributions. Additionally, by analyzing their usage, we can pinpoint critical genes for various software domains. These encompass common practices and foundational software genes, which can be harnessed to guide future software development, enhancing efficiency.

\subsubsection{Phenotype Analysis → Software Portrait.}

% Apart from the defective software genes we identified, there are more generalized software defects hidden in the commonly used third party software artifacts, such as code smells, poor maintainability, potential low code quality, etc. Although these software defects do not directly affect the regular functionality of software, they could still introduce potential threats to the long-terms maintenance of software. Therefore, it is still of great importance to properly identify these software defects and take countermeasures to enhance the long run of software.
% The objective of this task is to summarize these kinds of software defects in a comprehensive way, including not only factors of different dimensions on software assurance, but also proper and measurable metrics to appropriately reflect these factors to build up the profile for software genes. Besides the taxonomy, we also aim to propose a set of tool chains to properly identify and measure them in a well-organized way, and aggregate these measurement to standardized metrics.

Beyond the specific defective software genes we've pinpointed, more generalized software defects lurk within widely-used third-party software artifacts. These encompass code smells, potential maintainability issues, and indications of subpar code quality. While these defects might not immediately compromise software functionality, they can pose challenges for long-term software maintenance. Recognizing and addressing these defects is crucial for ensuring software longevity. Therefore, we aim to establish a comprehensive taxonomy of these software defects, and based on them, develop a suite of tools to systematically identify, measure, and aggregate these defects into standardized metrics.

\emph{\textbf{Software Assessment Taxonomy.}}
% First, we review existing academic papers and related tools from both academic and industrial, and retrieve the factors that currently researchers and developers care during their development. Specifically, we conclude the taxonomy of software defects from the following dimensions: 1) Software Quality, including the wrong and bad code smells in software which could lead to software defects during runtime; 2) Software Security, denoting the potential security threats identified as defects in software; 3) OSS compositions, representing the potential threats introduced by reused third-party artifacts, such as libraries and code snippets; 4) Maintainability, reflecting the maintenance of software components in different repositories, for instance, software genes from inactive repositories tend to have poor maintenance when issues happen; 5) Business Risk, annotating the commercial risks of software due to license and copyright issues hidden in reused third party artifacts. Based on these coarse-grained classifications, we aim to propose a comprehensive taxonomy to summarize these various software defects.
Initially, we surveyed existing academic literature and tools from both scholarly and industrial sources to identify factors that researchers and developers prioritize during development. From our findings, we've delineated a taxonomy of software defects across several dimensions:
1) Software Quality: This encompasses incorrect or suboptimal code patterns that might lead to runtime defects.
2) Software Security: This refers to potential vulnerabilities that can be exploited, compromising the software's integrity.
3) OSS Compositions: This highlights potential risks associated with the integration of third-party components, such as libraries or code snippets.
4) Maintainability: This dimension gauges the upkeep of software components across repositories. For instance, software genes from dormant repositories might be less reliable due to potential neglect.
5) Business Risk: This captures the commercial pitfalls linked to licensing and copyright concerns in reused third-party components.
With these broad categories in place, our objective is to craft a detailed taxonomy that encapsulates the diverse range of software defects.

\emph{\textbf{Software Defect Measurement.}}
% With the support of fine-grained taxonomy of software defects, we then develop the tool chains to properly measure these factors in comprehensive ways. Considering that not all factors can be perfectly detected by existing tools, we propose a combination strategy that aggregates the detection outcomes of different tools to reflect the performance of software genes on different dimensions. Specifically, different weights should be granted for different defects detected by these tools, for example, vulnerabilities should be more weighted that code smells as they bring more severe consequences and could be more probably to be utilized. 
Leveraging our detailed taxonomy of software defects, we advance toolchains to accurately gauge these factors. Recognizing that current tools might not capture all factors flawlessly, we advocate a combined approach, amalgamating results from various tools to represent software genes' performance across dimensions. Importantly, we'll assign varying weights to defects based on their severity and potential impact. For instance, vulnerabilities, due to their graver implications, would be weighted more heavily than code smells.

\subsubsection{Evolution Analysis → Software Evolution Analysis.}

Software genes, akin to biological counterparts, evolve over time. Understanding the evolution of the open-source ecosystem is pivotal to grasp its growth, dynamics, and adaptability, which could offer insights into trends and challenges, guiding strategic decisions and policy formulation. Such insights are crucial for improving the quality, security, and inclusivity of open-source software, ensuring its continued global influence.

\begin{figure*}[h]
\includegraphics[width=1\textwidth]{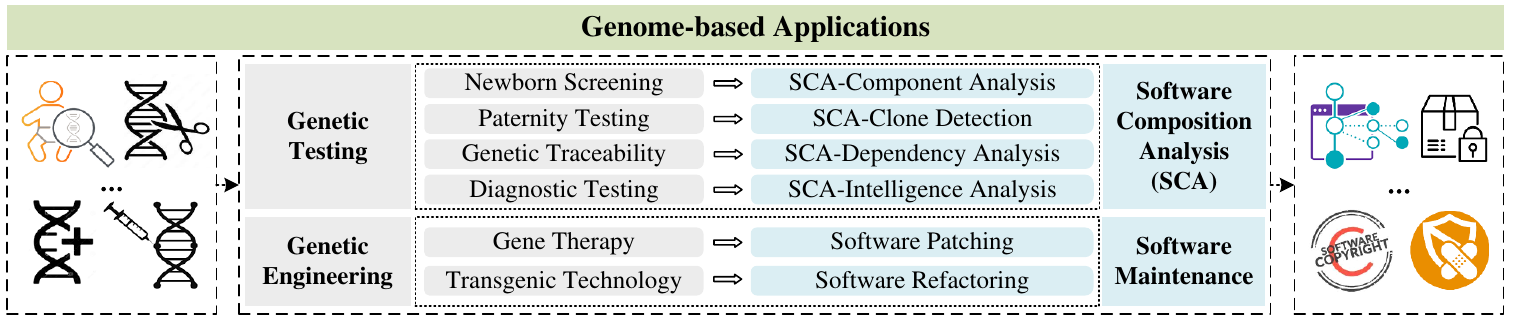}
\caption{Genome-based applications of software genome project}
\label{fig:applications}
%\vspace{-1em}
\end{figure*}

We aim to track the dynamic shifts of software genes in open-source ecosystems, pinpointing the nature, causes, and trends of changes during evolution. We aim to discern shifts in software development practices over time. For instance, prevalent coding practices might evolve with technological advancements. For core software genes, we examine how developers adapt these genes, observing additions, removals, or replacements to meet varying needs. Such observations can inform automated software development in the future. For domain-specific genes, we monitor their cross-domain introductions and interactions, aiding in understanding technical stacks and guiding component selection in software design. It's also vital to observe the rise and fall of support for various domains, ensuring the sustainability of key infrastructures, especially when faced with maintenance challenges. Through this, we can anticipate technological evolution in software development and offer timely recommendations.

%% file: outline/6_Genome_based_Appilcations.tex
\section{Genome-based Applications}
In this phase, we aim to translate the insights gained from our software genome analysis into practical real-world applications. 
Specifically, we model genetic testing and genetic engineering to develop predictive models to enhance software composition analysis (SCA) \cite{sca} and software maintenance.
Figure \ref{fig:applications} shows the details.

\subsection{Software Composition Analysis}
In the context of SCA, achieving our goals involves approaching genetic testing from four distinct perspectives. 
Component analysis \cite{rothenberger2003strategies} is akin to the concept of newborn screening \cite{wilcken2008newborn}, clone detection \cite{wang2023comparison} resembles paternity testing \cite{ma2006paternity}, dependency analysis \cite{sharma2009dependency} is comparable to genetic traceability \cite{dalvit2007genetic}, and vulnerability analysis \cite{zhao2023software} serves a role similar to diagnostic testing \cite{wallach2007interpretation}. 
This multi-faceted approach allows us to comprehensively assess and understand the genetic makeup of software components for security and reliability.

\subsubsection{Newborn Screening → SCA-Component Analysis.}
In this step, the code of the new software project is compared to our collected software genome. 
First, we design some advanced pattern recognition techniques to identify specific code structures and other structural characteristics that align with those in our collected software genome. 
Then, we conduct semantic analysis to understand the functional aspects of the code. 
This involves assessing how the code segments in the new project perform similar functions to those in our collected software genome.
Furthermore, we also identify whether the code segments in the new project correspond to specific versions of known software genes. 
After ensuring the version, we use some probabilistic matching algorithms to calculate the likelihood of a code segment in the new project being a match to a known gene. 
This helps in minimizing false positives and enhancing the precision of identification. 
By conducting these detailed processes, we ensure not only the precise identification of software components but also the preservation of their functional integrity and the comprehensive documentation of their characteristics. 

\subsubsection{Paternity Testing → SCA-Clone Detection.}
In this step, we design some novel code clone detection techniques which can scale to billion lines of code analysis. 
We will compute the code similarity between two code segments and compare it against a predefined threshold. 
Code segments that surpass the threshold are confirmed as code clones. 
These are instances where the code exhibits substantial commonality, much like confirming paternity when genetic markers closely match. 
Code clones are categorized based on the level of similarity and variations. 
This categorization provides insights into the heritage of the code and the extent of shared ancestry, similar to classifying genetic relationships. 
Each identified code clone is associated with metadata, including the location of code clones, their respective origins (source files or repositories), metadata about the authors or teams involved, and information regarding the extent of similarity. 
By executing these detailed process, we can systematically identify, assess, categorize, and document code clones within our software genome. 

\subsubsection{Genetic Traceability → SCA-Dependency Analysis.}
In this step, we conduct in-depth analysis of the dependencies to assess their characteristics, which is akin to evaluating the genetic traits or attributes inherited from ancestors. 
Specifically, we create a comprehensive lineage chart that depicts how dependencies are inherited within the software project and understand the parent-child relationships between components, which is akin to tracking genetic inheritance from one generation to the next.
Based on the lineage chart, we can identify the paths of inheritance, noting how specific dependencies are passed down to different parts of the project. 
In this way, we can recognize the dependencies that serve as ancestors, passing on their characteristics to descendant components.
During software traceability analysis, we will maintain detailed records including the attributes of each dependency, the name, version, functionality, and any identified issues. 
This documentation serves as a genetic trace archive, providing a historical record of dependency characteristics and their impact on the project.

\subsubsection{Diagnostic Testing → SCA-Intelligence Analysis.}
In this step, we conduct a detailed intelligence examination of each identified software component, which is akin to a comprehensive health checkup for humans.
As aforementioned, our intelligence genome consists of vulnerabilities, malicious code, and sensitive code.
Therefore, we complete the examination by analyzing all of them.
Specifically, vulnerability analysis is carried out by cross-referencing detected components with collected vulnerabilities in our intelligence genome to identify known weaknesses and exposures. 
Malicious code analysis involves utilizing advanced code scanning tools to compare the components with malicious codes in our intelligence genome to detect potential malware, viruses, or suspicious patterns that might indicate malicious intent.
Sensitive code analysis focuses on identifying sections of the code responsible for handling sensitive data, like personal, financial, or confidential information. 
It is completed by examine the sensitive code pattern retrieved by our intelligence genome analysis.
Our comprehensive analysis approach can help community make informed decisions regarding software security and risk management.

\subsection{Software Maintenance}
In the realm of software maintenance \cite{bennett2000software}, our objectives are met through a multifaceted approach that draws parallels to genetic engineering. 
Software patching \cite{dadzie2005understanding} can be likened to the concept of gene therapy \cite{verma2005gene}, while software refactoring \cite{mens2004survey} can be compared to transgenic technology \cite{yamamoto2001ons}. 
They collectively allow us to address and enhance the health and performance of software systems, much like genetic engineering aims to improve the traits and characteristics of organisms.

\subsubsection{Gene Therapy → Software Patching.}
In this step, we aim to address the potential threats introduced by the software intelligence genome to achieve the most beneficial holistic outcome. 
Typically, there are three major solutions for handling potential threats: adopting patches, altering component versions, and replacing artifacts.
It is crucial to appropriately combining these three solutions to mitigate various software threats. 
Moreover, considering that not all identified software threats bring equal risks, our subsequent thrust is towards devising a context-aware prioritization mechanism. 
For example, security vulnerabilities should be differently handled as they are usually only able to be triggered with different conditions, and the potential risks they could cause also significantly vary. 
Meanwhile, fixing one potential threat could inadvertently introduce new issues such as introducing new security threats and introducing artifacts with poor maintenance and code quality. 
We will circumvent these secondary challenges during patching to assure a seamless and comprehensive resolution.

\subsubsection{Transgenic Technology → Software Refactoring.}
In this step, our goal is to provide refactoring recommendations for open-source software based on the outcomes of software genome analysis.
Software refactoring is a systematic process that improves the quality, readability, and maintainability of existing code without altering its external behavior. 
To achieve the goal, we can identify code smells to strategically address issues related to code duplication, complexity, and overall design. 
Next, by creating unit tests, selecting appropriate refactoring techniques, and maintaining consistency, our code can be transformed into a more robust and efficient form. 
Moreover, we also employ some other methods like version control, code reviews, and performance monitoring to ensure that the refactoring process is reliable and does not introduce new defects. 
By conducting the previous process, we can contribute to the ongoing health and sustainability of software projects, make them more adaptable to evolving requirements and easier to understand for developers, thus enhancing their long-term success.

%% file: outline/7_Open_Source_Governance.tex
\section{Open Source Governance}
% The objective of this WP is to establish robust open-source governance. Drawing upon the comprehensive analysis outcomes of WP 1 through WP 5, our primary goal is to operationalize a suite of pivotal initiatives that encompass real-time intelligence prewarning, secured package manager, compatible semantic versioning, and license compliance. These initiatives collectively constitute the foundation upon which a well-structured and strategically governed open-source ecosystem will be built.

% Moreover, the software genome analysis also brings insights for open source governance, intended to guide the evolution of the open source ecosystem toward healthier and more standardized directions. In this section, we highlight countermeasures from the perspectives of policy and guidelines for open source governance.

Furthermore, unlike the natural evolution of biological genes, we, software practitioners, are truly manipulating the software genes by creating various software. In this case, the analysis of software genome also provides valuable insights into open source governance, enabling it to guide the evolution of the open source ecosystem towards healthier and more standardized directions. This section focuses on countermeasures from the perspectives of policy guidelines aimed at ensuring effective open-source governance.

\subsection{OSS Governance Policies}
% To facilitate the substainable development of open source ecosystem, some policies are compulsorily to be followed so that the reuse of open source artifacts would not cause catastrophic consequence. Some standards and policies, such as standards of semantic versioning~\cite{semver},  Software Bill of Material (SBOM)~\cite{SBOM} (i.e., SPDX, CycloneDX, and CPE standards~\cite{sbom2}) and open source license compliance and security assurance (i.e., ISO/IEC 5230~\cite{ISOIEC5230}, ISO/IEC DIS 18974~\cite{ISOIEC18974}). Our goal is to promote the widespread adoption of these standards through insights obtained from software genome analysis.
To ensure sustainable development in the open-source ecosystem, adherence to crucial standards and policies is imperative. For instance, as the rise of universal awareness on supply chain security, standards on different aspects of open source governance such as semantic versioning~\cite{semver}, Software Bill of Material (SBOM)~\cite{SBOM}, and open-source license compliance (ISO/IEC 5230~\cite{ISOIEC5230}) have been constantly raised. 
Our goal is to promote the widespread adoption of these standards through insights obtained from software genome analysis, and in this section, we take license compliance and semantic versioning as examples to elaborate our proposal. 

\subsubsection{License Compliance.}

% Reusing code fragments or third-party libraries from existing repositories has significantly accelerated the modern software development, while it also places potential legal risks of violating license compatibility due to lack of legal expertise. In this case, as one of the most crucial defects, it is vital to properly examining and managing the usage of licensed artifacts, which could largely decrease the legal risks on reusing existing coding features in modern software development.
% Our objective is to propose a complementary and comprehensive metamodel of software licenses, and based on it, collect and interpret licenses of existing code fragments, especially the code features in software genome, in a proper manner. Such an in-depth interpretation of software licenses could significantly enhance the risk management of license-related intelligence across the open-source ecosystem.

The widespread use of code fragments and third-party libraries in modern software development accelerates progress but introduces legal risks due to license compatibility issues. To mitigate these risks, a thorough examination and management of licensed artifacts, especially within the software genome, is crucial. Our goal is to introduce a comprehensive software license metamodel and leverage it to collect and interpret licenses systematically, enhancing risk management of license-related intelligence across the open-source ecosystem.

\emph{\textbf{License Metamodel.}}
% First, starting with the well-known software licenses, such as SPDX licenses and FOSS licenses, we plan to conduct an in-depth analysis to interpret the rights and obligations defined in these popular software licenses. Specifically, though existing researches have labelled existing software licenses with terms of rights and obligations, they are still with the coarse-grained interpretations which fails to capture more fine-grained requirements and scenarios. We aim to collect existing interpretations that are in different labelling system, and regulate them in a more standardized way by differential analysis. Apart from existing term classifications, we will also step forward to propose a comprehensive metamodel for more fine-grained license modelling. 
% Commencing with well-established software licenses such as SPDX licenses and FOSS licenses, our initiative involves a comprehensive analysis to discern the intricacies of rights and obligations stipulated in these widely recognized licenses. While previous research endeavors have categorized these licenses in terms of rights and obligations, their interpretations remain somewhat coarse-grained, lacking the granularity required to encompass more nuanced requirements and scenarios. Our approach aims to amalgamate existing interpretations, which may be formulated under different labeling systems, and standardize them through a meticulous differential analysis.
Our undertaking involves a meticulous analysis of well-known software licenses like SPDX and FOSS licenses~\cite{SPDXLicenseList, FreeSoftwareFoundation}, aiming to refine interpretations of rights and obligations. Despite prior efforts categorizing these licenses, existing interpretations lack granularity, necessitating our approach of amalgamating diverse interpretations under different labeling systems and standardizing them differentially. Subsequently, we propose a fine-grained license metamodel to address limitations in current interpretations.

\emph{\textbf{License Conflict Modelling.}}
% Next, we will also propose the license incompatibility model for code reuse based on our fine-grained license modelling. Specifically, there could be different types of license conflicts for different reasons. For instance, 1) Attitude conflicts. Reusing code fragments that are under restrictive licenses could be risky as obligations required could be easily ignored by project licenses (especially permissive licenses), which could be easily neglected by maintainers. 2) Conditional Conflicts. Some licenses only grant specific rights under specific conditions, while these conditions could be neglected if only term-level conflicts are taken into consideration. 3) Copyleft Conflicts. Copyleft licenses naturally require the granted rights to be inherited by downstream users of licensed code fragments, otherwise illegality could be introduced in downstream user projects. In this case, based on the license metamodel, we will further conclude a complementary and comprehensive license conflict model to guide further risk management on license compatibility.
Next, we are formulating a comprehensive license incompatibility model for code reuse. This model recognizes various conflicts, including attitude conflicts where reusing restrictive licenses poses risks, conditional conflicts arising from specific grant conditions, and copyleft conflicts mandating the inheritance of granted rights. Leveraging the metamodel, our overarching objective is to guide robust risk management practices related to license compatibility in code reuse scenarios.

\emph{\textbf{Customized License Identification.}}
% Moreover, besides popular software licenses, there are still massive customized software licenses in the open source ecosystem. In this case, based on the license modelling, we aim to empower the ability of recognizing the diverse customized licenses and interpret them into our metamodel. Specifically, starting with the license texts of existing well-interpreted licenses, we feed them, as well as the structured modelling of these licenses, into large language models (LLMs) to guide the classification of licenses. We aim to reuse the ability of LLMs on interpreting license texts and guide them to modelling these licenses based on our own license metamodel. Based on this, we are able to automate the interpretation of the massive customized licenses and convert them into a structured model, with which, further incompatibility analysis could be further conducted, and more fine-grained risk management could be properly carried out.
In addition to widely recognized software licenses, the open-source ecosystem encompasses a multitude of customized licenses. To enhance our capacity for recognizing and interpreting this diverse array of customized licenses within our metamodel, we employ a methodology that involves leveraging large language models (LLMs). Beginning with the license texts of well-interpreted licenses, we utilize LLMs to guide the classification of licenses, employing their adeptness in interpreting license texts. Subsequently, this interpretative ability is harnessed to model these licenses based on our proprietary license metamodel. This approach facilitates the automated interpretation of numerous customized licenses, transforming them into a structured model. This structured model, in turn, enables further incompatibility analysis and the implementation of more fine-grained risk management practices.

% In the dynamic landscape of the open-source ecosystem, where widely recognized licenses coexist with a myriad of customized counterparts, we seek to bolster our metamodel's recognition capabilities. Employing large language models (LLMs), our methodology commences with well-interpreted licenses. LLMs classify licenses, harnessing their adept text interpretation to seamlessly integrate them into our framework. This automated process not only expedites the nuanced interpretation of customized licenses but also lays the groundwork for comprehensive incompatibility analysis and the implementation of refined risk management practices.

\subsubsection{Compatible Semantic Versioning.}

% Open source software is usually maintained and released with certain agreement on its compatibility by following the semantic versioning, which plays a crucial role in software development, ensuring that updates and changes to software are systematically and clearly communicated to developers and end-users alike. Generally, it follows a structured format of version numbering and is essential for tracking software changes, managing dependencies, and maintaining system stability. However, in practice, inconsistencies and deviations in adhering to semantic versioning can lead to confusion, dependency errors, and system breakdowns. Therefore, the intricate landscape of software evolution, characterized by frequent updates and iterative enhancements, necessitates a stringent, standardized approach to adhere semantic versioning.

Open-source software typically adheres to a vital practice known as semantic versioning, playing a pivotal role in software development. This practice ensures systematic and clear communication of updates and changes to both developers and end-users. Following a structured version numbering format, semantic versioning is essential for tracking software changes, managing dependencies, and upholding system stability. However, practical implementation sometimes sees inconsistencies and deviations, leading to confusion, dependency errors, and system breakdowns. Consequently, the complex realm of software evolution, marked by frequent updates and iterative enhancements, mandates a stringent and standardized adherence to semantic versioning.

% Therefore, this task aims to propose proper solutions to detecting violations and improving the adherence of software releases with semantic versioning. Specifically, we aim to achieve ecosystem-wide mitigation of incompatible releases by three steps. 1) Compatibility detection tool: existing researches focus on detecting software incompatible changes by check API signatures, and fail to detecting the impact of changes on intra-procedural functionalities that happen in the internal changes of API implementation, leaving behavioural incompatibilities undetectable. In this case, we first integrate compatibility detection with finer-grained behavioural change detection to enhance compatibility tools. 2) Compatible release check: from the perspective of library maintainers, we are able to tell whether an artifact to be released are compatible with previous releases based on our compatibility tool, and based on this, we set up pipelines to monitor the daily release in major ecosystems, and trigger warnings when incompatible non-major releases happen. 3) Compatible upgrades: from the perspective of users, dependency updates are non-trivial to downstream users for the purpose of including new features and fixing necessary bugs. However, conducting such dependency upgrades could introduce compatible issues that could compromise user software. We aim to propose corresponding tools to help users finding released versions that are compatible with their deployed ones and manage the dependency configurations with proper flexibility on dependency ranges.

Thus, the primary objective of this task is to proffer comprehensive solutions aimed at detecting violations and strengthening adherence to semantic versioning within software releases. The proposed strategy unfolds through a tripartite approach, seeking to attain ecosystem-wide mitigation of incompatible releases. Firstly, enhancing compatibility detection tools by integrating finer-grained behavioral change detection. This innovation addresses a gap in current research, ensuring a more nuanced understanding of intra-procedural functionalities within API implementations. Secondly, from the vantage point of library maintainers, enabling the assessment of artifact compatibility with previous releases. This capability, facilitated by our compatibility tool, is coupled with the establishment of monitoring pipelines for daily releases in major ecosystems, issuing warnings in the event of incompatible non-major releases. Lastly, considering users, identifying released versions compatible with their deployed ones during dependency updates. Furthermore, our goal includes the provision of flexible tools for managing dependency configurations, thereby ensuring a seamless and user-friendly approach to dependency upgrades while preserving software stability.

% This set of compatible semantic versioning tools aim to provide solutions on managing software dependencies in a flexible but compatible way, which could benefit the management of technical lags and software defect mitigation from the perspective of both upstream maintenance and downstream usage, developers and users can have clear, unambiguous insights into the changes embedded in each software release, fostering informed decision-making, seamless integration, and system stability.

\subsection{OSS Governance Guidelines}

Apart from standards, we also aim to promote the sustainability of open source ecosystems by introducing guidelines to the community, such as toolkits on identifying fundamental artifacts and guiding community supports based on software genome analysis, monitoring, alarming, and remediating critical threats based on disease genome analysis, measuring and crediting contributions by incentives, to facilitate the substainability open source communities. 
In this section, we also take two examples, i.e., 1) pre-warning and monitoring to critical threats and 2) secured package managers to isolate untrustworthy artifacts, to elaborate our countermeasures.

\subsubsection{Real-time Intelligence Prewarning.}

The rapidly expanding open-source ecosystem, with its numerous contributors, faces increased security vulnerabilities. The simultaneous development of many projects raises the chances of unresolved bugs and security issues. A comprehensive intelligence monitoring system is crucial to address these challenges, enabling stakeholders to quickly identify and resolve issues, thus maintaining software integrity and strengthening the ecosystem's resilience.

Our main goal is to continuously monitor repositories, analyze code changes, and track forums and contributor activities for potential threats. The system includes an advanced warning feature to alert developers, users, and administrators about new vulnerabilities or unusual behaviors. Central to this system is a well-curated open-source software repository and a sophisticated security intelligence framework, designed for rapid and accurate threat detection. By integrating these elements, we aim to create a dynamic toolkit that not only enhances software analysis efficiency but also reinforces software ecosystems against new risks and threats.

\emph{\textbf{Lightweight and Fast SCA.}}
Identifying security threats in open-source repositories and components is a complex task. To tackle this, we propose two efficient solutions. First, at the source-code level, we suggest POM-based and code-clone-based methods for lightweight SCA to boost performance. Moreover, as many software artifacts are released as binaries in software ecosystems, inspecting them usually involves complex program analysis to decompile binary code and extract control flow graphs for component identification. This process is time-consuming, often taking hours for a single binary, and becomes impractical for large systems with thousands of files. Additionally, the lack of detailed features from program analysis hinders accurate detection using syntactic attributes.
To overcome these issues, our two-pronged strategy involves: 1) refining the open-source dataset by removing duplicates, thereby reducing noise from internal dependencies; and 2) extracting strings from projects to serve as lightweight features. We cluster these strings based on macros in the project source code, utilizing the likelihood of their co-occurrence in compiled binaries. An adaptive rating system is employed to assess the relevance of an open-source project based on the matching percentage of string clusters. Our approach strives for precise SCA outcomes while maintaining remarkable performance, particularly in analyzing large-scale projects.

\emph{\textbf{Security Threat Prediction and Confirmation.}}
After completing SCA, we obtain a detailed bill of materials for the analyzed projects, which lays the groundwork for the next crucial step: identifying security threats in open-source packages. This task is achieved by utilizing a security threats database to establish a strong link between identified threats and their respective projects. To increase accuracy and minimize false positives, we employ a multi-faceted strategy.
Firstly, we design pipelines to monitor open-source security intelligence, capturing and processing new reports and news about software defects. Additionally, we use program analysis in tandem with a sophisticated large language model (LLM). This LLM understands the semantics of potential vulnerability in function candidates. As for vulnerabilities in binaries, we extract functional features to deduce the collective purpose of binary instruction groups. The LLM then generates high-level abstractions of the functions' behaviors, enabling a detailed evaluation of their integrity, and particularly, identifying any malicious intent.
When malicious behavior is detected, the vulnerability is confirmed, and a thorough explanation of its cause is provided. This comprehensive approach ensures not only the accurate detection of vulnerabilities but also offers deep insights into the nature of the security issues identified.

\subsubsection{Secured Package Manager.}

The extensive reuse of third-party artifacts from public repositories introduces various software supply chain threats, including those from dependency imports and code fragment reuse. To mitigate these risks, this project aims to create a secured package manager infrastructure that ensures safe reuse of third-party artifacts for downstream users.

This secured package manager comprises six key components:
1) Secured Registry: A three-tier artifact registry hosting reusable artifacts. It includes a patched registry for artifacts with known defects, an audited registry for monitoring commonly used artifacts, and a public registry for accessing public repositories.
2) Secured Advisory: An advisory database documenting the defects of open-source artifacts. It leverages open-source intelligence to monitor user application supply chains.
3) Client: A user interface for managing dependencies on third-party artifacts, allowing for secure management and pre-publishing checks.
4) Backend Engine: A server-side engine optimizing the reuse of third-party artifacts, focusing on prioritizing dependency resolution to avoid defects and controlling the publishing and withdrawal of artifacts.
5) Auditing Engine: A cluster of engines analyzing open-source artifacts to generate comprehensive profiles of user-required artifacts, guiding user behaviors like publishing, withdrawing, and importing.
6) Configuration Language: A domain-specific language for configuring the reuse of third-party artifacts, catering to both existing packages and commonly reused code fragments. This allows users to set their dependency requirements, including thresholds and preferences for resolution and remediation.

The secured package manager is designed to be secure and lightweight, facilitating the publication, maintenance, distribution, and integration of third-party artifact reuse. It enables the effective use of the Software Genome's open-source intelligence system, enhancing open-source governance and surveillance capabilities.

%% file: outline/8_Conclusion.tex
\section{Conclusion}
Drawing inspiration from the Human Genome Project, this paper presents a detailed approach to comprehending software composition by likening it to human genetic formation. This analogy is used to develop precise strategies for enhancing software maintenance and guiding the evolution of the software ecosystem. Specifically, the paper delves into the granularity of code snippets to understand software formation and to identify the diversity and capacity of coding knowledge within the ecosystem. The insights gained from this software genome analysis can also shed light on the software ecosystem's evolution and may offer valuable reference points for understanding other evolving subjects.